# Whether relative respiration in trees can be constant: a discussion of a scaling hypothesis

V.L. Gavrikov

**Abstract**

Respiration measurements of whole tree plants have been reported that give evidence that the relative per volume/mass unit respiration decreases with increase of tree body size. In this study, based on the available data published a question was explored if the relative per area unit respiration in trees can be a constant, independent of the surface area size. There is a definite gap in the published data when the allometric studies of tree body structure do not intercept with studies on trees respiration. Thus the question was studied with the help of indirect comparison between various data. The comparison showed that the scaling exponents, volume vs. surface area and respiration vs. stem volume, are slightly larger than they should be for the hypothesis of the relative respiration constancy to hold. The data studied give evidence that the relative per area unit respiration slightly increases with the increase in tree surface area. Possible explanations of the relationship include a different distribution of metabolically active parts of stem and higher nitrogen content in larger trees. Also, the published datasets might include large fast growing trees, which imply that larger trees grow faster and hence have higher per unit surface area growth respiration. A crucial experiment is required in which the respiration measurements were performed for the same data as the measurements of scaling between stem volume and surface area.

**Introduction**

Respiration is known to be one of the most fundamental processes in living bodies. The interest in respiration in trees originates both from the wish to understand the basics of life and the need to estimate the global role of forests in climate change.

Direct measurements of trees respiration on a whole plant level are rather scarce because of obvious technical difficulties. Nevertheless some recent reports provided important data on the matter of whole plant respiration and the scaling of it across species and plant sizes. Reich et al. (2006) reported of measurements of about individual 500 plants belonging to 43 species that were both laboratory and field-grown specimens. In these data, the trees were represented by saplings and seedlings. As a result, Reich et al. (2006) argued that the whole-plant respiration rate scales approximately isometrically with total plant mass, which means that the power exponent is varying about unity.

Cheng et al. (2010) gave the evidence of respiration rates in larger trees. They showed that the aboveground respiration rates scaled as 0,82-power of the aboveground biomass. Mori et al. (2010) conducted an extensive research of direct measuring of respiration in 271 whole plants, from seedlings to large trees. The authors established that the scaling allometric exponent varied continuously from unity for smallest plants to 3/4 in larger saplings and trees.

Obviously, the total respiration of a plant body grows as the body's total size increases. But the relative respiration per unit of body size may show at least two distinct behaviors with the body growth. Let $R$ stand for the total respiration, $\gamma$ for the scaling exponent and $V$ for the plant body volume. (For the sake of consistence, I will use the plant volume $V$ as a measure of the total body size assuming a good relationship between body volume and body mass). So, the scaling relationship is given by

$$R \propto V^{\gamma}. \qquad (1)$$

If the scaling exponent $\gamma$ is equal to unity then per volume unit respiration should be constant, i.e. independent of $V$, since $R/V \propto V^0$. If however the scaling exponent $\gamma$ is less than unity then per volume unit respiration cannot be a constant but should be a decreasing function of $V$:

$$\frac{R}{V} \propto V^{\gamma-1}.$$

It is widely understood that for larger trees the relative, per volume unit, respiration decreases with the growth of the relevant body size, the volume. The biological grounds for this have been also clearly discussed. While the bodies of smaller plants are metabolically active in the whole volume, low-active stem wood constitutes most of the biomass of larger trees (Pruyn et al., 2005; Mori et al., 2010).

On the other hand, the same logic can be applied in respect to another measure of tree size, the stem surface area. The respiration of the tree stem is largely located in the thin sheath of inner bark (Pruyn et al., 2002, 2005). Unlike stem volume, an increase in the bole surface area is always the relevant increase in living metabolically active tissues. It is therefore natural to hypothesize that the relative per area unit respiration can be a constant independent of the amount of the surface area.

Mathematically, the hypothesis is expressed as follows. Because the relationship between stem volume $V$ and stem surface area $S$ may be expressed through a scaling exponent $\beta$ as

$$V \propto S^{\beta} \qquad (2)$$

then substituting (2) in (1) one gets the expression for the total respiration as

$$R \propto S^{\gamma\beta}.$$

Respectively, the relative per area unit respiration can be given as

$$\frac{R}{S} \propto S^{\gamma\beta-1}. \qquad (3)$$

It is obvious that the scaling exponent $\gamma\beta-1$ in (3) may be equal to zero, and for this the relation between the scaling exponents $\beta$ and $\gamma$ should be as follows

$$\gamma = \frac{1}{\beta}. \qquad (4)$$

The aim the study was to test the hypothesis if the relative per unit area respiration in larger trees can be a constant, i.e. independent of the amount of bole surface area, on the basis of available public data. Formally, the hypothesis is presented in the relationship (4).

**Materials and methods**

There is a definite gap in the published data. Studies focused on tree respiration do not provide values of the scaling exponents between stem volume and stem surface area. Studies aimed at estimating of the morphological scaling exponents do not deal with measuring of tree respiration.

The values for γ were taken from the sources cited above (Reich et al., 2006; Cheng et al., 2010; Mori et al., 2010). Additionally, the dataset published by Cheng et al. (2010) was partly recalculated. The dataset contains DBHs, heights, log-transformed respiration and log-transformed mass parameters for a number of species; among them are two conifers, Pinus tabulaeformis Carr. and Pinus massoniana Lamb. For these two conifers, a two-way volume equation (Inoue, Kurokawa, 2001) was applied to get bole volumes. The bole surface areas were estimated through a cone surface formula. Using of the two measures, the scaling exponent β was calculated by fitting of the combined pine data by a power function. Independently, the log-transformed respiration and mass measures were fitted by a linear function to get the exponent γ for the same combined pine data.

The values for β were taken from a study by Inoue and Nishizono (personal communication) on relationship between stem volume and stem surface area in Japanese cedar and Japanese cypress forest stands. The values for β were also estimated from the datasets of levels-of-growing-stock studies in Douglas-fir (Marshall, Curtis, 2001; Curtis et al., 2009). For control plots in the datasets, bole volumes were calculated by dividing of stand volume by the number of living trees for every age available. From the same tables, the data of mean quadratic diameter and mean stand heights were taken to estimate the mean bole surface areas through the cone formula. Then the mean volume was fitted against mean bole surface area by a power function to get estimations of the β power exponent.

All the fittings were performed by means of STATISTICA 6 software using an ordinary least squares approach.

**Results and discussion**

The found in the literature and estimated values of γ and β scaling exponents are summarized in the table. In most of the cases if one of the two exponents is available the other is not. That is why the unknown exponent was calculated in accordance with the hypothesis (4) and given in the table in parentheses.

Table. Reported and estimated values of scaling exponents γ and β.

| Reported and estimated values of scaling exponents | | References |
|---|---|---|
| β | γ | |
| (1 to 1,33)* | from 1 to 0,75 | Mori et al., 2010 |
| (1,22) | 0,82 | Cheng et al., 2010 |
| 1,568±0,032** | 0,7429±0,032** | Cheng et al., 2010 (recalculated) |
| (1) | 1 | Reich et al., 2006 |
| from 1,35 to 1,79 | (0,74 to 0,56) | Inoue, Nishizono, Japanese cedar |
| from 1,35 to 1,63 | (0,74 to 0,61) | Inoue, Nishizono, Japanese cypress |
| 1,569±0,006** | (0,64) | Hoskins experiment, Douglas-fir |
| 1,487±0,007** | (0,67) | Iron Creek experiment, Douglas-fir |

\* in parentheses, estimated through (4) values are given

\*\* fitted values ± std. error

According to the hypothesis (4) if the relative per area unit respiration is independent of the surface area then the scaling exponents γ and β should exactly compensate each other so that their product is equal to unity. An examination of the data in the table shows that the measured values of the scaling exponent β tend to be slightly bigger that those expected through (4). For example, the multi-species study by Mori et al. (2010) gives the minimal value of γ as 0,75 which through (4) corresponds to the maximal value of β of ≈1,33. The study by Cheng et al.

(2010) suggest the value 0,82 for γ which gives ≈1,22 for β. The measurements by Inoue and Nishizono (personal communication) give the values of β from 1,35 and larger.

On the other hand, the values for γ tend to be some larger than those expected through (4). For example, the maximum γ values estimated through (4) for Inoue and Nishizono and Hoskins and Iron Creek data amount 0,74. The minimal γ value measured in Reich et al. (2006), Cheng et al. (2010) and Mori et al. (2010) is 0,75. All the comparisons mean that the product of γ and β should be slightly bigger than unity.

In the only case, data by Cheng et al. (2010), it was possible to estimate the scaling exponents γ and β for the same dataset (see in table 'recalculated'). Multiplication of γ and β for the data subset gives 1,568×0,7429 ≈ 1,16. Thus the relation (3) for the case should be as $R/S = S^{0,16}$ which means that the relative per area unit respiration should slightly grow with the increase in the total surface area.

The inference may look to some extent counterintuitive. In fact, if the very properties of stem surface in the process of growth would remain the same then there were not sufficient causes for the relative per area unit respiration to alter. The data shown in the table give evidence that in some, rather peculiar, cases the scaling exponents γ and β may satisfy (4) and therefore provide certain stability of the per unit area respiration. Nevertheless it looks more likely that in general the scaling exponents γ and β do not satisfy (4) and they are expected to produce the increase of relative per unit area respiration.

A couple of hypothesis can be suggested to explain the increase of the per unit area respiration. The allometric scaling concept provide useful generalizations of tree body structure but does not take account of the complexity of tree stem in terms of physiological variation of its different parts. It is known that respiration vary in stems and branches of different diameters and branching orders (Bosc et al., 2003). It has been also shown that respiration in tree tissues strongly linked to the content of nitrogen in them (Reich et al., 2008). It is not unlikely that the distribution of variously active surfaces within stems of larger trees is different from that in smaller trees. Pruyn et al. (2002) found that sapwood of older trees had higher respiratory potential than sapwood of younger trees if the outer-bark surface area of stems was used as a basis for comparing respiratory potential.

Another consideration deals with the overall non-linearity of tree growth. It is widely known that a typical tree has an S-shaped growth curve, both in linear and volumetric terms. If one considers a dataset with no mature and over-matured trees then for this particular dataset larger trees will on the average grow faster. Because growth respiration is a sufficient part of the overall respiration the growth curve non-linearity means that larger trees would show higher respiration on both absolute and relative per area unit basis.

As a conclusion it should be admitted that these explanations remain theoretical speculations until a crucial experiment is performed. The crucial experiment should include measurements of both scaling exponents, volume vs. surface area and respiration vs. volume/mass, on the same dataset.

**References**

Bosc, A., De Grandcourt, A., & Loustau, D. (2003). Variability of stem and branch maintenance respiration in a Pinus pinaster tree. Tree Physiology, 23(4), 227-236.

Cheng, D. L., Li, T., Zhong, Q. L., & Wang, G. X. (2010). Scaling relationship between tree respiration rates and biomass. Biology letters, 6 (5), 715–717, doi:10.1098/rsbl.2010.0070.

Inoue A., Kurokawa Y. (2001). Theoretical Derivation of a Two-way Volume Equation in Coniferous Species. The journal of the Japanese forestry society, 83(2), 130-134.

Curtis, R. O. Levels-of-growing-stock Cooperative Study in Douglas-fir: Report No. 19–the Iron Creek Study, 1966-2006 / R. O. Curtis, D. D. Marshall et al.– United States Department of Agriculture, Forest Service, Pacific Northwest Research Station, 2009.– 78 pp.

Marshall, D. D. Levels-of-growing-stock cooperative study in Douglas-fir: report no. 15-Hoskins: 1963-1998 / D. D. Marshall, R. O. Curtis.– United States Department of Agriculture, Forest Service, 2001.–80 pp.

Mori, S., Yamaji, K., Ishida, A., Prokushkin, S. G., Masyagina, O. V., Hagihara, A., ... & Umari, M. (2010). Mixed-power scaling of whole-plant respiration from seedlings to giant trees. Proceedings of the National Academy of Sciences, 107(4), 1447–1451.

Pruyn, M. L., Gartner, B. L., & Harmon, M. E. (2002). Respiratory potential in sapwood of old versus young ponderosa pine trees in the Pacific Northwest. Tree Physiology, 22(2-3), 105–116.

Pruyn, M. L., Gartner, B. L., & Harmon, M. E. (2005). Storage versus substrate limitation to bole respiratory potential in two coniferous tree species of contrasting sapwood width. Journal of experimental botany, 56(420), 2637-2649.

Reich, P. B., Tjoelker, M. G., Machado, J. L., & Oleksyn, J. (2006). Universal scaling of respiratory metabolism, size and nitrogen in plants. Nature, 439(7075), 457-461.

Reich, P. B., Tjoelker, M. G., Pregitzer, K. S., Wright, I. J., Oleksyn, J., & Machado, J. L. (2008). Scaling of respiration to nitrogen in leaves, stems and roots of higher land plants. Ecology letters, 11(8), 793–801.